\documentclass[aps,preprint, pe]{revtex4}
\usepackage{amsmath}
\begin{document}

\title{A unified approach to the derivation of work theorems for equilibrium and steady-state,
classical and quantum Hamiltonian systems}

\author{M. F. Gelin}
\author{D. S. Kosov}

\affiliation{Department of Chemistry and Biochemistry, University of Maryland, College Park, MD  20742}

\begin{abstract}  
We present a unified and simple method for deriving work theorems 
for classical and quantum Hamiltonian systems, both under
equilibrium conditions and in a steady state. 
Throughout the paper, we adopt the partitioning of the total Hamiltonian
into the system part, the bath part, and their coupling.
We rederive many equalities which are available in the literature 
and obtain a number of new equalities for nonequilibrium classical and quantum systems.  
Our results can be useful
for determining partition functions and (generalized)
free energies through simulations or measurements performed on
nonequilibrium systems. 
\end{abstract}
\maketitle

\section{Introduction }

The fluctuation theorems allow us to rigorously relate equilibrium
ensemble properties of a dynamical system with its evolution under
nonequilibrium conditions, beyond the domain of validity of the linear
response theory.\cite{kuz81a,kuz81b,spo99,eva02,rit03,kur07,wil07,bro07,roe07} 
Many  recent results, concepts, and ideas are stemmed from the early landmark 
work by Bochkov and Kuzovlev.\cite{kuz81a,kuz81b}   

Among the fluctuation theorems, 
the Jarzynski equality (or, equivalently, the nonequilibrium work theorem) occupies a
remarkable place.\cite{jar97,jar04,jar07} This equality connects the nonequilibrium work performed
on a system with the ratio of the equilibrium system's free energies.
Over the past ten years, the Jarzynski equality has been extended to non-Markovian stochastic processes,\cite{sei07} 
to quantum systems \cite{kuz81a,roe07,yuk00,muk03,muk04,muk06,mae04,mon05,nie05,han07,nol07} and to systems
coupled to different (non-Hamiltonian) thermostats. \cite{eva02,wil07,sch04,del06,cue06}
Several important results for dissipative systems in the steady state
have also been obtained. \cite{sas01,sei05,sei05a,jar06,qia07,coh07,der07} 

The fluctuation theorems and   
Jarzynski equality have been proven for a Hamiltonian system coupled to Hamiltonian heat 
bath(s), see Refs.\cite{jar00,ver03} and Ref.\cite{jar04}, respectively.  
The present paper is aimed at presenting a unified and simple method
for generating various work theorems  for such systems. We
consider both classical and quantum systems, which can initially be
prepared either under equilibrium conditions or in a steady state.
Within our approach, we rederive many equalities which are available in the literature 
and obtain a number of new results for nonequilibrium classical and quantum systems.  
Our expressions can be considered as mathematical identities, since  
the fulfillment of the Liouville theorem is required only.  They 
can be useful for determining various (equilibrium or steady state) partition functions
and (generalized) free energies through simulations and/or measurements
performed on nonequilibrium systems. 
The nonequilibrium partition functions can be used in much the same manner as their equilibrium counterparts. Indeed, our steady-state distributions have the generic form $\rho_{ne} =1/Z_{ne} \exp(\sum_n g_n G_n)$, where 
$Z_{ne}$ is the non-equilibrium partition function, $g_n$   are certain c-numbers, and $G_n$  are the corresponding operators. If we differentiate the logarithm of the partition function   with respect to the parameter $g_n$, we obtain the expectation value of the operator $G_n$. Doing so, we can get expectation values of the steady-state energy, entropy, particle number, etc. Caution should be exercised in quantum case, however, because operators  may not commute with each other.

\section{Classical systems}

Let $H(\Gamma,t)$ be the Hamiltonian (which is allowed to
be explicitly time-dependent), $\Gamma$ be the corresponding phase
variables, and $A(\Gamma,t)$, $B(\Gamma,t)$ be certain non-pathological
functions of the phase variables and time. Then we can write down
the identity\begin{equation}
A(\Gamma_{0},0)A(\Gamma_{0},0)^{-1}B(\Gamma_{t},t)=B(\Gamma_{t},t),\label{AB}\end{equation}
$\Gamma_{0}$ and $\Gamma_{t}$ being the values of the phase variables
at the time moments $0$ and $t$. We can integrate Eq. (\ref{AB}) over
$\Gamma_{0}$ and make use of the fact that the motion of a Hamiltonian
system can be regarded as a canonical transformation, for which
the Liouville theorem holds: $d\Gamma_{0}$$=d\Gamma_{t}$. We thus
obtain \begin{equation}
\int d\Gamma_{0}A(\Gamma_{0},0)\left(A(\Gamma_{0},0)^{-1}B(\Gamma_{t},t)\right)=\int d\Gamma_{t}B(\Gamma_{t},t)\equiv B_{t}.\label{AB1}\end{equation}

If we assume that $A(\Gamma_{0},0)$ is positively defined and
normalized ($\int d\Gamma_{0}A(\Gamma_{0},0)=1$), we can consider
$A$ as the probability density, denote the averaging \begin{equation}
\int d\Gamma_{0}A(\Gamma_{0},0)...\equiv\left\langle ...\right\rangle_{A}\label{av}\end{equation}
and rewrite Eq. (\ref{AB1}) as \begin{equation}
\left\langle A(\Gamma_{0},0)^{-1}B(\Gamma_{t},t)\right\rangle _{A}=B_{t}.\label{AB2}\end{equation}

\subsection{Systems at equilibrium }

The proof of Eq. (\ref{AB2}) given above is very similar to that of the
Jarzynski equality for Hamiltonian systems.\cite{jar04} However,
Eq. (\ref{AB2}) contains the Jarzynski equality and much more. Indeed,
let both $A$ and $B$ be the instantaneous Gibbs distributions:
\begin{equation}
A(\Gamma_{0},t)=\rho_{0}(\Gamma_{0})=Z_{0}^{-1}\exp(-\beta H(\Gamma_{0},0)),\,\,\, B(\Gamma_{t},t)=\rho_{t}(\Gamma_{t})=Z_{t}^{-1}\exp(-\beta H(\Gamma_{t},t)).\label{ABeq}\end{equation}
Here $H(\Gamma_{t},t)$ is a time-dependent Hamiltonian, \begin{equation}
Z_{0}=\int d\Gamma_{0}\exp(-\beta H(\Gamma_{0},0)),\,\,\, Z_{t}=\int d\Gamma_{t}\exp(-\beta H(\Gamma_{t},t))\label{Z}\end{equation}
are the corresponding partition functions and $\beta=1/(k_{B}T)$,
$k_{B}$ being the Boltzmann constant and  $T$ being the temperature. Plugging Eqs. (\ref{ABeq})
into our starting formula (\ref{AB2}) we get\begin{equation}
\left\langle \exp(-\beta(H(\Gamma_{t},t)-H(\Gamma_{0},0)))\right\rangle _{\rho_{0}}=Z_{t}/Z_{0}.\label{Ja}\end{equation}
The time derivative of  any function $C(\Gamma_{t},t)$ is determined by the following expression:
\begin{equation}
\frac{d}{dt}C(\Gamma_{t},t)=\frac{\partial}{\partial t}C(\Gamma_{t},t)+\{ C(\Gamma_{t},t),H(\Gamma_{t},t)\},\label{Poiss}\end{equation}
$\{...\}$ being the Poisson bracket. Thus\begin{equation}
C(\Gamma_{t},t)-C(\Gamma_{\tau},\tau)\equiv\int_{\tau}^{t}dC(\Gamma_{t'},t')=\int_{\tau}^{t}dt'\frac{\partial}{\partial t'}C(\Gamma_{t'},t')+\int_{\tau}^{t}dt'\dot{\Gamma}_{t'}\frac{\partial}{\partial\Gamma_{t'}}C(\Gamma_{t'},t').\label{ChainC}\end{equation}

We partition  the total Hamiltonian into the system Hamiltonian,
the bath Hamiltonian, and their coupling:
\begin{equation}
H(\Gamma_{t},t)=H_{S}(x_{t},t)+H_{B}(X_{t})+H_{SB}(x_{t},X_{t}).\label{SB}\end{equation}
Here $x_{t}$ and $X_{t}$ are the phase variables specifying the
system and the bath, and the system Hamiltonian only is allowed to
be explicitly time-dependent. Plugging $H(\Gamma_{t},t)$ (\ref{SB})
into identity (\ref{ChainC}) and making use of the fact that
$\{ H(\Gamma_{t},t),H(\Gamma_{t},t)\}\equiv0$, we can write\begin{equation}
H(\Gamma_{t},t)-H(\Gamma_{0},0)=\int_{0}^{t}dt'\frac{\partial}{\partial t'}H(\Gamma_{t'},t')=\int_{0}^{t}dt'\frac{\partial}{\partial t'}H_{S}(x_{t'},t')\equiv W,\label{Ja1}\end{equation}
$W$ being the work performed on the system. Thus Eqs. (\ref{Ja})
and (\ref{Ja1}) yield the Jarzynski formula \cite{jar04,foot2}\begin{equation}
\left\langle \exp(-\beta W)\right\rangle _{\rho_{0}}=Z_{t}/Z_{0}.\label{Ja0}\end{equation}

Eq. (\ref{AB2}) can be applied to more complicated situations. Let
us assume that the system and the bath are initially prepared at different
temperatures $T_{S}$ ($\beta_{S}=1/(k_{B}T_{S})$) and $T$, respectively.
We can take $A$ and $B$ to be the corresponding nonequilibrium distributions\[
A(\Gamma_{0},0)=\rho_{\beta0}(\Gamma_{0})=Z_{\beta0}^{-1}\exp(-\beta H(\Gamma_{0},0)-(\beta_{S}-\beta)H_{S}(x_{0},0)),\]
\begin{equation}
B(\Gamma_{t},t)=\rho_{\beta t}(\Gamma_{t})=Z_{\beta t}^{-1}\exp(-\beta H(\Gamma_{t},t)-(\beta_{S}-\beta)H_{S}(x_{t},t)),\label{ABT}\end{equation}
$Z_{\beta0}$ and $Z_{\beta t}$ being the corresponding partition
functions. Inserting these formulas into Eq. (\ref{AB2}), we obtain\begin{equation}
\left\langle \exp(-\beta_{S}W-(\beta_{S}-\beta)Q\right\rangle _{\rho_{\beta0}}=Z_{\beta t}/Z_{\beta0}.\label{JaT}\end{equation}
Here the work $W$ is explicitly defined via Eq. (\ref{Ja1}) and
\begin{equation}
Q\equiv\int_{0}^{t}dt'\dot{x}_{t'}\frac{\partial}{\partial x_{t'}}H_{S}(x_{t'},t')
\label{Q}
\end{equation}
can be regarded as a heat, which is transfered to the system.  This definition of $Q$ can be understood based on the following consideration. The energy of the system can be changed by performing the work $W$ on the system or by supplying heat $Q$ to the system:
\begin{equation}
H_S(x_t, t) - H_S(x_0, 0) = \int_{0}^{t}dt'\frac{\partial}{\partial t'}H_{S}(x_{t'},t') + \int_{0}^{t}dt'\dot{x}_{t'}\frac{\partial}{\partial x_{t'}}H_{S}(x_{t'},t').
\end{equation}
Since the first term in this equation is the work (\ref{Ja1}), then the second term can be associated with the heat absorbed by the system.

Eq. (\ref{JaT}) can be considered
as the generalized Jarzynski equality. It means that (nonequilibrium)
entropy can be changed by making some work and/or exchanging some
heat. If we assume that the Hamiltonians $H$ and $H_{S}$ do not
depend on time explicitly, then Eq. (\ref{JaT}) reduces to \begin{equation}
\left\langle \exp(-(\beta_{S}-\beta)Q)\right\rangle _{\rho_{\beta0}}=1,\label{JaT1}\end{equation}
which is the identity derived in Ref.\cite{jar04a}

\subsection{Systems in a steady state}

To derive the steady-state distribution, we can also apply the
procedure of the external driving of the molecular Hamiltonian.\cite{zub,gra,her93,bok05,mor04,and06,fuj07,foot4}
Namely, we assume that the system-bath interaction is 
switched on adiabatically, so that the total time-dependent Hamiltonian reads as
\begin{equation}
H(\Gamma_{t},t)=H_{S}(x_{t})+H_{B}(X_{t})+\exp(\varepsilon t)H_{SB}(x_{t},X_{t})\label{H(t)}\end{equation}
($\varepsilon$ is a positive infinitesimal). At a certain moment
in the past $t=\tau\ll-1/\varepsilon$ our {}``system'' and {}``bath''
do not interact and are prepared at the temperature $T$ according to
the grand canonical Gibbs distribution \begin{equation}
\rho_{\tau}=Z_{\tau}^{-1}\exp(-\beta(H(\Gamma_{\tau},\tau)-Y(\Gamma_{\tau})).\label{Req0C}\end{equation}
Here \begin{equation}
H(\Gamma_{\tau},\tau)=H_{S}(x_{\tau})+H_{B}(X_{\tau}),\,\,\, Y(\Gamma_{\tau})=\mu_{S}N_{S}(x_{\tau})+\mu_{B}N_{B}(X_{\tau}),\label{Req01}\end{equation}
$\mu_{\alpha}$ are the chemical potentials for $S$ and $B$. $N_{S}(x_{t})=1$ if the spatial coordinate in the phase point $x_{t}$  belongs to the volume $V_{S}$ occupied by the system and zero otherwise; similarly for $N_{B}(X_{t})$. 

Preparation of the ensemble according to distribution (\ref{Req0C}) means that, initially, the system $S$ and 
the bath $B$ were in equilibrium with different heat baths. At the moment $\tau$, $S$ and $B$ are decoupled from their baths, and the $S-B$ interaction is gradually switching on.     
As a result of this incipient interaction, the
systems can exchange their particles and energies with each other, so that the
steady state is established at $t=0$.\cite{foot1} To arrive at the desirable
steady-state distribution, we can proceed as follows. First, we apply
the thermodynamic limit to the bath degrees of freedom, so that $N_{B}\rightarrow\infty$,
$V_{B}\rightarrow\infty$, $N_{B}/V_{B}\rightarrow\mathrm{const}$
($N_{B}$ is the number of bath particles and $V_{B}$ is the volumes occupied by the bath). Second,
we tend $\varepsilon $ to zero (i.e., $\tau\rightarrow-\infty$) and propagate the initial equilibrium
distribution (\ref{Req0C}) from $t=\tau$ to $t=0$ with the time-dependent
Hamiltonian (\ref{H(t)}). Third, we change from the phase variables
$\Gamma$ (at $t=\tau=-\infty$) to $\Gamma_{0}=\Gamma_{0}(\Gamma,t=0)$.
Then the non-equilibrium steady-state distribution at $t=0$ can be
written as follows \cite{zub,gra,her93,bok05,mor04,and06,fuj07}
\begin{equation}
\rho_{s0}=Z_{s0}^{-1}\exp(-\beta(H(\Gamma_{0},0)-Y(\Gamma_{0}))).\label{Rne}\end{equation}
Here $Z_{s0}$ is the steady-state partition function \cite{foot3}
and $Y(\Gamma_{0})$ obeys the identity $\{ Y(\Gamma_{0}),H(\Gamma_{0},0)\}=0$.
There are several equivalent forms of $Y(\Gamma_{0})$. \cite{zub,gra,her93,bok05,mor04,and06,fuj07}
For the further discussion, the most elucidating is the following
expression, which can be obtained if we substitute function $Y$ for $C$  in Eq.(\ref{ChainC}), set $t=0$ and use the fact that the total number of particle is conserved ($d/dt (N_S +N_B)=0$):
\begin{equation}
Y(\Gamma_{0})=Y(\Gamma_{\tau})+\Delta_{\mu}\overline{J}(\Gamma_{0}).\label{Y}\end{equation}
Here $\Delta_{\mu}=\mu_{B}-\mu_{S}$ and \begin{equation}
\overline{J}(\Gamma_{0})=\int_{\tau}^{0}dt'\dot{\Gamma}_{t'}\frac{\partial}{\partial\Gamma_{t'}}Y(\Gamma_{t'})\label{Jss}\end{equation}
is the time-integrated current density. 
The explicit form (\ref{Y}) of $Y(\Gamma_{0})$ makes evident a profound distinction between the equilibrium and steady-state preparation. If the chemical potentials of the system and the bath are the same, $\Delta_{\mu}=0$, then the combined $S-B$ system will end up with an equilibrium distribution. Otherwise, the steady-state distribution establishes. It supports the steady-state currents, which are absent in equilibrium.  
It should be noted that Eq. (\ref{Rne}) does not rely upon any sort of perturbation or linear response theory, and thus describes the steady-state distribution far from equilibrium. Furthermore, Eqs. (\ref{Rne})-(\ref{Jss})
deliver an explicit formula for the steady-state distribution in terms of Hamiltonian (\ref{H(t)}).   

We are in the position now to derive the work theorem
for the steady-state systems. If we assume that $A=\rho_{\tau}$ and
$B=\rho_{s0}$, we get then\begin{equation}
\left\langle \exp(-\beta(W_{SB}-\Delta Y)\right\rangle _{\rho_{\tau}}=Z_{s0}/Z_{\tau}.\label{SSonset}\end{equation}
Explicitly,\begin{equation}
W_{SB}\equiv H(\Gamma_{0},0)-H(\Gamma_{\tau},\tau)=\varepsilon\int_{\tau}^{0}dt'\exp(\varepsilon t')H_{SB}(x_{t'},X_{t'}),\label{Wsb}\end{equation}
and \begin{equation}
\Delta Y\equiv Y(\Gamma_{0})-Y(\Gamma_{\tau})=\Delta_{\mu}\overline{J}.\label{DY}\end{equation}
Eq. (\ref{SSonset}) allows us to follow the energy exchange during
the onset of the steady state. 
By its definition (\ref{Wsb}), $W_{SB}$ looks similar to the 
nonequilibrium work in the standard Jarzynski equality (\ref{Ja0}). However, the quantity cannot be associated with the work performed on (by) the system. $W_{SB}$ can be regarded as  the time-averaged value of the system-bath coupling
$H_{SB}$. This follows immediately from the Abel's theorem, \cite{zub}
which states that the identity \begin{equation}
\lim_{\varepsilon\rightarrow+0}\varepsilon\int_{-\infty}^{0}dt'e^{\varepsilon t'}f(t')=\lim_{t\rightarrow\infty}\frac{1}{t}\int_{-t}^{0}dt'f(t'),\label{Ab}\end{equation}
holds for any function (operator) $f(t)$. On
the other hand, $\Delta Y$ is proportional to the time-integrated current.
Its presence in Eq. (\ref{SSonset}) is peculiar to the steady-state distribution,
since $\Delta Y$ vanishes in equilibrium. The value of $\Delta Y$ equals 
the  additional energy we have to spend for establishing
the steady-state distribution.

An interesting result is obtained if we introduce the distribution 
\begin{equation}
\rho_{H}=Z_{H}^{-1}\exp(-\beta(H(\Gamma_{\tau},0)-Y(\Gamma_{\tau}))),\label{RneH}\end{equation}
$Z_{H}$ being the corresponding partition function. Eq. (\ref{RneH}) assumes  
that the system and the bath are coupled all the time, but their chemical potentials are kept different. 
$\rho_{H}$ is neither an equilibrium nor the steady-state distribution. However, it is a 
perfectly legitimate mathematical object to consider. If we take $A=\rho_{s0}$ (\ref{Rne}) and $B=\rho_{H}$ (\ref{RneH}), we get 
\begin{equation}
\left\langle \exp(-\beta\Delta_{\mu}\overline{J}(\Gamma_{0}))\right\rangle _{\rho_{s0}}=Z_{H}/Z_{s0}.\label{Jtok}\end{equation}
Such a result can be obtained through the fluctuation theorems for currents,\cite{kuz81a,eva02,wil07,gas07a,gas07b,muk07,mae08}
and a similar formula has been derived in Ref. \cite{mor04} in the context 
of the shear flow steady-state thermodynamics.   The only difference between these results and ours is as follows: $Z_{H}/Z_{s0}\neq1$, in general. If the system-bath coupling is weak, however, we can write that  $Z_{H}/Z_{s0}=1+O(H_{SB})$ and the ratio of the partition functions equals one in the leading order.

We assume now that our $S+B$
system is prepared at $t=0$ in the steady-state distribution (\ref{Rne}).
Then, we switch the external field on at $t=0$, so that the driven
system Hamiltonian is explicitly described by Eq. (\ref{SB}) at $t>0$.
Under the influence of such a Hamiltonian, the steady-state distribution
(\ref{Rne}) will evolve into \begin{equation}
\rho_{st}=Z_{st}^{-1}\exp(-\beta(H(\Gamma_{t},t)-Y(\Gamma_{t}))),\label{Rnet}\end{equation}
$t>0$. If we take $A=\rho_{s0}$ (Eq. (\ref{Rne})) and $B=\rho_{st}$
(Eq. (\ref{Rnet})), we obtain 
\begin{equation}
\left\langle \exp(-\beta(W-(Y(\Gamma_{t})-Y(\Gamma_{0}))\right\rangle _{\rho_{s0}}=Z_{st}/Z_{s0}.\label{JaY1}\end{equation}
Here the work $W$ is explicitly defined via Eq. (\ref{Ja1}) and
\begin{equation}
\Delta Y_{s}\equiv Y(\Gamma_{t})-Y(\Gamma_{0})=\int_{0}^{t}dt'\dot{\Gamma}_{t'}\frac{\partial}{\partial\Gamma_{t'}}Y(\Gamma_{t'}).\label{DY}\end{equation}
Again, the additional term $\Delta Y_{s}$ enters
Eq. (\ref{DY}) as compared to its Jarzynski counterpart (\ref{Ja0}),
manifesting thereby the necessity of additional energy expenses in the steady state. 
$\Delta Y_{s}$ is always positive since it is the product of the current 
$\overline{J}$ and the chemical potential difference $\Delta_{\mu}$ and 
they always have the same sign.

\subsection{Additional useful equalities and fluctuation theorems}

(i). The (information) entropy associated with any nonequilibrium distribution
can be defined as \begin{equation}
S_{a}\equiv-k_{B}\ln\rho_{a}.\label{Seq}\end{equation}
If we take $A=\rho_{a}$ and $B=\rho_{b}$ (the subscripts $a$
and $b$ correspond to any probability density function introduced
above), then Eq. (\ref{AB2}) tells us that\begin{equation}
\left\langle \exp((S_{a}-S_{b})/k_{B})\right\rangle _{a}=1.\label{Ent}\end{equation}
This expression is thus very general and  independent
of a particular form of the nonequilibrium distribution.\cite{kuz81a,eva02,wil07} It states that the path average
of the exponential of the entropy production equals unity. For stochastic systems, a
similar result has been proven in Refs. \cite{sei05,cro99} It should be noted that, in general, 
$S_{a}-S_{b}$ in Eq. (\ref{Ent}) is the total system+bath entropy production,       
while the papers \cite{sei05,cro99} show that Eq. (\ref{Ent}) holds for the 
entropy production of the system only, provided that the system dynamics is described by the Markovian
master equation. It can be argued that if the bath is infinite then its entropy does not change and  $S_{a}-S_{b}$ is 
referred to the system. Furthermore, if $A$ and $B$ are distributions  (\ref{ABeq}) or (\ref{ABT}), then $S_{a}-S_{b}$ is 
rigorously determined by the system operators only. 

(ii). If we multiply Eq. (\ref{AB}) by $\delta(\Gamma-\Gamma_{t})$
and integrate over $\Gamma_{t}$, we obtain the identity \begin{equation}
\int d\Gamma_{0}A(\Gamma_{0},0)\delta(\Gamma-\Gamma_{t})\left(A(\Gamma_{0},0)^{-1}B(\Gamma_{t},t)\right)=B(\Gamma,t).\label{FK}\end{equation}
If we take $A$ and $B$ to be the instantaneous Gibbs distributions (\ref{ABeq}),
then we recover the expressions derived in Refs.
\cite{cro00,sza01} If $A$ and $B$ are certain nonequilibrium distributions,
we arrive at the result derived in \cite{jar05} for the system describing
via an overdamped Langevin equation. 

(iii). Let 
\begin{equation}
C(\Gamma_{t},t)=\Psi(D(\Gamma_{t},t)), \,\,\, D(\Gamma_{t},t)\equiv A(\Gamma_{0},0)^{-1}B(\Gamma_{t},t), \label{C}\end{equation}
$\Psi(D)$ being a certain function. We also introduce the inverse function, so that $\Psi^{-1}(C)=D$. If we multiply Eq. (\ref{AB}) by $\delta(w+C(\Gamma_{t},t))$ ($w$ being a parameter), integrate it over $\Gamma_{t}$, and use the notation (\ref{av}), then 
we obtain the identity \begin{equation}
\Psi^{-1}(-w)\left\langle\delta(w+C(\Gamma_{t},t))\right\rangle _{A}=\left\langle\delta(w+C(\Gamma_{t},t))\right\rangle _{B}\label{C1}\end{equation}
which can be coined as the generalized Crooks transient fluctuation theorem. 
If we take $A$ and $B$ to be the instantaneous Gibbs distributions (\ref{ABeq}),
and let $C$ be the "minus" forward work performed on the system (\ref{Ja1}),
\begin{equation}
C(\Gamma_{t},t)=W(\Gamma_{t},t)=-W(\Gamma_{0},t)=H(\Gamma_{0},0)-H(\Gamma_{t},t)=\frac{1}{\beta}\ln \left(D(\Gamma_{t},t)\frac{Z_{t}}{Z_{0}} \right),\label{C2}\end{equation}
then we obtain the Crooks transient  fluctuation theorem\cite{cro99,cro00}
\begin{equation}
\left\langle\delta(w-W(\Gamma_{0},t))\right\rangle _{\rho_{0}}=\left\langle\delta(w+W(\Gamma_{t},t))\right\rangle _{\rho_{t}}\exp(\beta w)Z_{t}/Z_{0}.\label{Cr}\end{equation}

Furthermore, let $A$ and $B$ be  \textit{any} (equilibrium or not) distributions evolving into each other along the forward and time-reversed trajectories, respectively. If we assume that $C=\Psi(D)=k_{B}\ln D$ and adopt definition (\ref{Seq}) for the entropy, then Eq. (\ref{C1}) yields the Crooks equality for the entropy production $\Delta S(\Gamma_{0},t)=-C(\Gamma_{0},t)$: 
\begin{equation} 
\left\langle\delta(w-\Delta S (\Gamma_{0},t))\right\rangle _{\rho_{0}}=\left\langle\delta(w+\Delta S(\Gamma_{t},t))\right\rangle _{\rho_{t}}\exp(w/k_{B}).\label{Cr1}\end{equation}
Thus, as has been shown in Ref. \cite{cro99},  Eq. (\ref{Cr1}) is valid if we start from any, not necessary equilibrium, distribution. 
Eq. (\ref{Cr1})  has been derived in Ref. \cite{jar00} for an externally driven Hamiltonian system coupled to several 
Hamiltonian heat reservoirs kept at different temperatures. This result is recovered if $A$ and $B$ are taken as nonequilibrium distributions (\ref{ABT}). 

(iv). We can generate complimentary work theorems by interchanging
$A$ and $B$ in Eq. (\ref{AB1}). For example, the so-obtained
analogue of Eq. (\ref{Jtok}) reads
\begin{equation}
\left\langle \exp(\beta\Delta_{\mu}\overline{J})\right\rangle _{\rho_{H}}=Z_{s0}/Z_{H}.\label{Jtok1}\end{equation}
Therefore,  \begin{equation}
\left\langle \exp(\beta \Delta_{\mu}\overline{J})\right\rangle _{\rho_{H}}\left\langle \exp(-\beta \Delta_{\mu}\overline{J})\right\rangle _{\rho_{s0}}=1.\label{Jtok2}\end{equation}

\section{Quantum systems}

Almost all results obtained in the previous section are immediately
transferable to the quantum case. In fact, we have to replace
all the functions by operators in the Heisenberg representation (hereafter,
the latter are marked by hats), Poisson brackets by commutators, and
integrations by taking traces. Thus, our ``generating expressions''
(\ref{AB1}) and (\ref{AB2}) transform into \begin{equation}
\mathrm{Tr}(\hat{A}(0)\left(\hat{A}(0)^{-1}\hat{B}(t)\right))\equiv\left\langle \hat{A}(0)^{-1}\hat{B}(t)\right\rangle_{A} =\mathrm{Tr}\hat{B}(t)\equiv B_{t},\label{AB1Q}\end{equation}

\begin{equation}
\left\langle \hat{A}(0)^{-1}\hat{B}(t)\right\rangle_{A} =B_{t}.\label{AB2Q}\end{equation}
Explicitly, the time dependence of any Heisenberg operator $\hat{C}$
is governed by the quantum analogue of the equation of motion (\ref{Poiss})
\begin{equation}
\frac{d}{dt}\hat{C}(t)=\frac{\partial}{\partial t}\hat{C}(t)+i[\hat{H}(t),\hat{C}(t)].\label{PoissQ}\end{equation}
$[...]$ is the commutator and we use the units with $\hbar=1$. Thus
\begin{equation}
\hat{C}(t)-\hat{C}(\tau)\equiv\int_{\tau}^{t}d\hat{C}(t')=\int_{\tau}^{t}dt'\frac{\partial}{\partial t'}\hat{C}(t')+\int_{\tau}^{t}dt'i[\hat{C}(t'),\hat{H}(t')]\label{ChainQ}.\end{equation}

\subsection{Systems at equilibrium }

We write the total Hamiltonian as a sum of the system Hamiltonian,
the bath Hamiltonian and their coupling and also allow for the system
Hamiltonian to be explicitly time dependent, \begin{equation}
\hat{H}(t)=\hat{H}_{S}(t)+\hat{H}_{B}+\hat{H}_{SB}.\label{SBQ1}\end{equation}
If both $A$ and $B$ are the instantaneous quantum canonical distributions\begin{equation}
\hat{A}(0)=\hat{\rho}_{0}=Z_{0}^{-1}\exp(-\beta\hat{H}(0)),\,\,\,\hat{B}(t)=\hat{\rho}_{t}=Z_{t}^{-1}\exp(-\beta\hat{H}(t)),\label{ABeqQ}\end{equation}
then we arrive at the expression\begin{equation}
\left\langle \exp(\beta\hat{H}(0))\exp(-\beta\hat{H}(t))\right\rangle_{\rho_{0}} =Z_{t}/Z_{0}.\label{YaQQ}\end{equation}
Now we can introduce the quantity \begin{equation}
\hat{W}=\hat{H}(t)-\hat{H}(0)=\int_{0}^{t}dt'\frac{\partial}{\partial t'}\hat{H}(t')=\int_{0}^{t}dt'\frac{\partial}{\partial t'}\hat{H}_{S}(t'),\label{WQ}\end{equation}
which is sometimes referred to as the work operator.\cite{yuk00,nie05,han07,nol07} If $[\hat{H}(t),\hat{H}(0)]=0$,
then Eq. (\ref{YaQQ}) transforms into the Jarzynski expression
\begin{equation}
\left\langle \exp(-\beta\hat{W})\right\rangle_{\rho_{0}} =Z_{t}/Z_{0},\label{YaQu}\end{equation}
If the Hamiltonians do not commute, then Eq. (\ref{YaQQ}) can be
rewritten as \begin{equation}
\left\langle \exp(-\beta(\hat{W}+\hat{\delta}_{W})\right\rangle_{\rho_{0}} =Z_{t}/Z_{0},\label{YaQu1}\end{equation}
where the quantum correction reads \begin{equation}
\hat{\delta}_{W}=-\frac{1}{\beta}\ln\left(\exp(\beta\hat{H}(0))\exp(-\beta\hat{H}(t))\right)+\hat{H}(0)-\hat{H}(t).\label{YaQu2}\end{equation}
This quantum correction arises  due to the fact that operators $\hat{H}(0)$ and  $\hat{H}(t)$ do not commute in (\ref{YaQQ}) and it is not associated with housekeeping 
heat.\cite{pan98,tas06}

Alternatively, Eq. (\ref{YaQQ}) can be recast into the form similar
to Eq. (\ref{YaQu}) even for non-commutative $\hat{H}(0)$ and $\hat{H}(t)$,
provided we introduce the chronological ordering operator $\mathcal{T}_{<}$:\cite{han07}
\begin{equation}
\mathcal{T}_{<}\left\langle \exp(-\beta \hat{W})\right\rangle_{\rho_{0}} =Z_{t}/Z_{0}.\label{Hron}\end{equation}
Here $\hat{W}$, due to Eq. (\ref{WQ}), is the work performed on the system $S$ only, in full analogy with the classical case. 
Eq. (\ref{Hron}) makes it clear that the path average of the time-ordered Heisenberg operator $\exp(-\beta \hat{W}(t))$ 
yields the ratio of the partition functions.  
See also \cite{kuz81a,yuk00,muk03,muk04,muk06,mae04,mon05,nie05,han07,nol07} for the derivation
of quantum Jarzynski equalities and discussion of various definitions
of the quantum work operator. 

If we consider the grand canonical ensemble, we arrive at a simple
generalization of Eqs. (\ref{YaQQ}) and (\ref{Hron}) provided we assume that 
the chemical potential $\mu(t)$ is  
externally driven. We can take 
\begin{equation}
\hat{A}(0)=\hat{\rho}_{N0}=Z_{N0}^{-1}\exp(-\beta(\hat{H}(0)-\mu(0)\hat{N})),\,\,\,\hat{B}(t)=\hat{\rho}_{Nt}=Z_{Nt}^{-1}\exp(-\beta(\hat{H}(0)-\mu(t)\hat{N})),\label{ABeqQN}\end{equation}
$\hat{N}$ being the number of particles operator, $Z_{N0}$ and
$Z_{Nt}$ being the corresponding partition functions. Eq. (\ref{AB2Q})
yields then \begin{equation}
\left\langle \exp(-\beta(\hat{W}+\hat{\delta}_{W}-\Delta\mu(t)\hat{N})\right\rangle_{\rho_{N0}} =Z_{Nt}/Z_{N0},\label{ABmu}\end{equation}
$\Delta\mu(t)=\mu(t)-\mu(0)$. If our
Hamiltonian does not depend on time explicitly, then Eq. (\ref{ABmu})
simplifies to \begin{equation}
\left\langle \exp(\beta\Delta\mu(t)\hat{N})\right\rangle_{\rho_{N0}} =Z_{Nt}/Z_{N0},\label{ABmu1}\end{equation}
irrespective of a particular driving protocol for $\Delta\mu(t)$. 

Straightforward is to derive quantum analogues of other results
obtained in Section IIA. For example, a quantum version of
Eq. (\ref{JaT}) reads \begin{equation}
\mathcal{T}_{<}\left\langle \exp(-\beta_{S}\hat{W}-(\beta_{S}-\beta)\hat{Q}\right\rangle _{\rho_{\beta0}}=Z_{\beta t}/Z_{\beta0}.\label{JaQ}\end{equation}
Here the work operator is defined via Eq. (\ref{WQ}) and the heat
operator is determined as \begin{equation}
\hat{Q}\equiv\int_{0}^{t}dt'i[\hat{H}(t'),\hat{H}_{S}(t')].\label{QQ}\end{equation}
Note the chronological ordering in Eq. (\ref{JaQ}), which takes care
of the fact that the operators do not commute, in general. We can also
generalize Eq. (\ref{JaQ}) by changing from the
canonical to grand canonical ensemble, thereby connecting the nonequilibrium
work, the transferred heat, and the particle exchange to the ratio
of two partition functions.

\subsection{Systems in a steady state}

Our consideration here parallels that of the classical systems in Sec. IIB.
We assume that the system-bath interaction is switched
on adiabatically, so that the total time-dependent Hamiltonian is written as \begin{equation}
\hat{H}=\hat{H}_{S}+\hat{H}_{B}+\exp(\varepsilon t)\hat{H}_{SB}\label{H(t)Q}\end{equation}
($\varepsilon$ is a positive infinitesimal). At a certain time moment
in the past $t=\tau\ll-1/\varepsilon$ our {}``system'' and {}``bath''
do not interact and are prepared at the temperature $T$ according to
the grand canonical Gibbs distribution \begin{equation}
\hat{\rho}_{\tau}=Z_{\tau}^{-1}\exp(-\beta(\hat{H}(\tau)-\hat{Y}(\tau))=Z_{\tau}^{-1}\exp(-\beta\hat{H}(\tau))\exp(\beta\hat{Y}(\tau)).\label{Req0}\end{equation}
Here $Z_{\tau}=\mathrm{Tr}(\hat{\rho}_{\tau})$, \begin{equation}
\hat{H}(\tau)\equiv\hat{H}_{S}+\hat{H}_{B},\,\,\,\hat{Y}(\tau)\equiv\mu_{S}\hat{N}_{S}+\mu_{B}\hat{N}_{B},\label{Y0}\end{equation}
$\mu_{a}$ are the chemical potentials for $S$ and $B$ and $\hat{N}_{a}$
are the corresponding number operators. Apparently, $[\hat{H}(\tau),\hat{Y}(\tau)]=0$.

If we apply the thermodynamic limit to the bath degrees of freedom
($N_{B}\rightarrow\infty$, $V_{B}\rightarrow\infty$, $N_{B}/V_{B}\rightarrow\mathrm{const}$)
and propagate the initial distribution $\hat{\rho}_{\tau}$ from $t=\tau$
to $t=0$, we arrive at the (nonequilibrium) steady-state distribution
at $t=0$:\cite{zub,gra,her93,bok05,and06,fuj07}
\begin{equation}
\hat{\rho}_{s0}=
Z_{s0}^{-1}\exp(-\beta(\hat{H}(0)-\hat{Y}(0))=Z_{s0}^{-1}\exp(-\beta\hat{H}(0))\exp(\beta\hat{Y}(0)).\label{RneQ}\end{equation}
Here $Z_{s0}=\mathrm{Tr}(\hat{\rho}_{s0})$ and, as in the classical
case, $[\hat{H}(0),\hat{Y}(0)]=0$. Explicitly, \begin{equation}
\hat{Y}(0)=\hat{Y}(\tau)+\Delta_{\mu}\hat{\overline{J}},\label{curr}\end{equation}
where the time-integrated current operator $\hat{\overline{J}}$ is determined
through the current operator $\hat{J}(t)=\hat{\dot{N}}_{B}(t)=i[\hat{H}(t),\hat{N}_{B}(t)]$
as 
\begin{equation}
\hat{\overline{J}}=\int_{\tau}^{0}dt'\hat{J}(t').\label{curr1}\end{equation}
Distribution (\ref{RneQ}) can be derived within the framework of the method of statistical operator by Zubarev \cite{zub} and (generalized version of the) maximum entropy principle by Jaynes.\cite{gra} The use of distribution (\ref{RneQ}) and the standard Keldysh Non-equilibrum Green's functions technique { yields  the same steady-state averages.\cite{her93,and06,fuj07}  The equivalence between Keldysh non-equilibrium Green's functions\cite{keldysh}
and Zubarev method of statitistical operator is discussed in the appendix.

Now we are in the position to derive quantum analogues of the expressions
presented in Sec. 2B. If we assume that $\hat{A}=\hat{\rho}_{\tau}$ and
$\hat{B}=\hat{\rho}_{s0}$, we obtain the expression\begin{equation}
\mathcal{T}_{<}\left\langle \exp(-\beta(\hat{W}_{SB}-\Delta\hat{Y})\right\rangle _{\rho_{\tau}}=Z_{s0}/Z_{\tau},\label{ssQ}\end{equation}
which imposes certain limits on the energy exchange during the onset
of the steady state. Explicitly,\begin{equation}
\hat{W}_{SB}\equiv\hat{H}(0)-\hat{H}(\tau)=\varepsilon\int_{\tau}^{0}dt'\exp(\varepsilon t')\hat{H}_{SB}(t'),\label{WsbQ}\end{equation}
and \begin{equation}
\Delta\hat{Y}\equiv\hat{Y}(0)-\hat{Y}(\tau)=\Delta_{\mu}\int_{\tau}^{0}dt'i[\hat{H}(t),\hat{N}_{B}(t)]=\Delta_{\mu}\hat{\overline{J}}.\label{DYQ}\end{equation}
As in the classical case, operator $\Delta\hat{Y}$ is responsible for the onset of the steady state, since 
$\Delta_{\mu}\hat{\overline{J}}=0$ vanishes at equilibrium. 
A quantum counterpart of Eq. (\ref{Jtok}) reads  
\begin{equation}
\mathcal{T}_{<} \left\langle \exp(-\beta\Delta_{\mu}\hat{\overline{J}}(\Gamma_{0}))\right\rangle _{\rho_{s0}}=Z_{H}/Z_{s0}.\label{JtokQ}\end{equation}

We can also write down a quantum extension of Eq. (\ref{DY}).
Let our $S+B$ system be prepared at $t=0$ according
to the steady-state distribution (\ref{RneQ}). Then, we switch the
external field on at $t=0$, so that the driven system Hamiltonian
is explicitly described by Eq. (\ref{SBQ1}) at $t>0$. Under the
influence of this Hamiltonian, the steady-state distribution (\ref{Rne})
will evolve into \begin{equation}
\hat{\rho}_{st}=Z_{st}^{-1}\exp(-\beta(\hat{H}(t)-\hat{Y}(t))),\label{RnetQ}\end{equation}
$t>0$. If we take $\hat{A}=\hat{\rho}_{s0}$ (Eq. (\ref{RneQ}))
and $\hat{B}=\hat{\rho}_{st}$ (Eq. (\ref{RnetQ})), we get\begin{equation}
\mathcal{T}_{<}\left\langle \exp(-\beta(\hat{W}-\Delta\hat{Y}_{s})\right\rangle _{\rho_{s0}}=Z_{st}/Z_{s0}.
\label{JaY1}\end{equation}
Here the quantum work $\hat{W}$ is explicitly defined via Eq. (\ref{WQ})
and \begin{equation}
\Delta\hat{Y}_{s}=\hat{Y}(t)-\hat{Y}(0)=\int_{0}^{t}dt'i[\hat{H}(t'),\hat{Y}(t')].\label{DYQ1}\end{equation}

If we define the quantum entropy operator 
as \cite{zub,gra}
\begin{equation}
\hat{S}_{a}\equiv-k_{B}\ln\hat{\rho}_{a},\label{SQ}\end{equation}
then the classical Eq. (\ref{Ent}) remains valid in the slightly
modified form: \begin{equation}
\left\langle \exp(\hat{S}_{a}/k_{B})\exp(-\hat{S}_{b}/k_{B})\right\rangle _{a}=1.\label{EntQ}\end{equation}
Here the subscripts $a$ and $b$ correspond to any density matrix
operator introduced above. For a quantum system described by the Markovian
master equation, a similar result has been derived in Ref. \cite{muk06} for the 
entropy production of the system. In Eq. (\ref{EntQ}), $\hat{S}_{a}$ and $\hat{S}_{b}$ 
are the entropy operators for system+bath and the (strong) system-bath coupling  
can affect the system entropy production due to the quantum entanglement. \cite{nie00}  By using the chronological ordering, Eq. (\ref{EntQ}) can be rewritten 
in the form similar to Eq. (\ref{Hron}).     

\subsection{Experimental or computational verification of quantum identities}

Within classical mechanics, it is conceptually straightforward  to measure or compute the evolution of a certain physical quantity along the trajectory. Thus, putting aside technical and computational difficulties,  verification and 
interpretation of the identities derived in Section II is primarily a matter of attributing a physical significance to the quantities $A$ and $B$. In quantum mechanics, it is not that clear how to measure or evaluate a certain physical quantity along the trajectory. An important question is therefore  how the 
quantum identities can be interpreted, experimentally verified,  or numerically tested
for nontrivial systems.  We discuss several possibilities in this section.

All our quantum identities can equivalently be recast into
the so-called two-time measurement form. \cite{muk03,muk04,muk06,mae04,mon05,nie05,han07,nol07}
To this end, let us
return back to our generating Eq. (\ref{AB1Q}). Without any loss of generality, the operators $A$ and $B$  can be written in the
exponential form
\begin{equation}
A(0)=\exp\{-\Lambda(0)\},\,\,\, B(t)=\exp\{-\Upsilon(t)\},
\label{ABexp}
\end{equation}
here we use the Schr\"{o}dinger representation for operators $A(0)$ and $B(t)$.
We assume that the operators $\Lambda(0)$ and $\Upsilon(t)$ are Hermitian.
Being the solutions of the eigenproblems for Hermitian operators 
\begin{equation}
\Lambda(0)\left|\lambda\right\rangle =E_{\lambda}\left|\lambda\right\rangle ,\,\,\,\Upsilon(t)\left|\upsilon_{t}\right\rangle =E_{\upsilon_{t}}\left|\upsilon_{t}\right\rangle \label{ABeig}
\end{equation}
($E_{\upsilon_{t}}$ and $\left|\upsilon_{t}\right\rangle $ depend on time parametrically) 
the eigenvectors are orthogonal
and complete. We also assume that these eigenvectors span the same Hilbert space.

We define the evolution operator $G(0,t)$ in the Liouville
space, which governs the time evolution of  our system according to \begin{equation}
G(0,t)\left|\lambda\right\rangle \left\langle \lambda\right|=\left|\lambda(t)\right\rangle \left\langle \lambda(t)\right|,\,\,\, G^{\dagger}(0,t)B(t)\equiv G^{\dagger}(0,t)\exp\{-\Upsilon(t)\}\equiv\widehat{B}(t).\label{Drive}\end{equation}
$G$ is unitary if the system dynamics is Hamiltonian, but it may
be not if we consider the dissipative system dynamics. Then Eq. (\ref{AB1Q})  can be rewritten in the following equivalent form:
\begin{equation}
\left\langle \hat{A}(0)^{-1}\hat{B}(t)\right\rangle_{A}=A_{0}\sum_{\lambda,\upsilon_{t}}\rho_{\lambda}\left|\left\langle \lambda(t)\right|\left.\upsilon_{t}\right\rangle \right|^{2}\exp\{-(E_{\upsilon_{t}}-E_{\lambda})\}\equiv A_{0}\left\langle \exp\{-\Delta_{E}\}\right\rangle=B_{t}.
\label{TwoTime}\end{equation}
Here \begin{equation}
\rho_{\lambda}=A_{0}^{-1}\exp\{-E_{\lambda}\},\,\,\, A_{0}=\mathrm{{\textstyle Tr}}\{ A(0)\}=\sum_{\lambda}\exp\{-E_{\lambda}\},\,\,\,\Delta_{E}\equiv E_{\upsilon_{t}}-E_{\lambda}.\label{Defin}\end{equation}
Eq. (\ref{TwoTime}) can be interpreted in terms of
the two-time measurement scheme. The first measurement at
$t=0$ selects an eigenfunction $\left|\lambda\right\rangle $ of
operator $\Lambda(0)$ in the Schr\"{o}dinger  representation. The second measurement at
time $t$ selects an eigenfunction $\left|\upsilon_{t}\right\rangle $
of operator $\Upsilon(t)$ also in Schr\"{o}dinger representation. The factor $\left|\left\langle \lambda(t)\right|\left.\upsilon_{t}\right\rangle \right|^{2}$ gives
us the probability of the system's evolution from $\left|\lambda\right\rangle $
to $\left|\upsilon_{t}\right\rangle $. If we repeat the the procedure
many times, we yield the mean value of $\left\langle \exp\{-\Delta_{E}\}\right\rangle$,
which is obtained by averaging over initial conditions and summation
over final conditions. Thus, Eqs. (\ref{TwoTime}) and (\ref{Defin}) present an explicit
measurement protocol, which (at least in principle) can be applied
to test all our quantum equalities derived in Sections 3a and 3b. Of course, the meaning and interpretation
of each equality depends on a particular form of (Hermitian) operators
$A$ and $B$ (or $\Lambda$ and $\Upsilon$), and a proper
interpretation is not a trivial task. \cite{footN} 

In several important cases, operators $\Lambda$ and $\Upsilon$ consist
of sums of two or more operators. If the later operators commute (as,
e.g., $\hat{H}$ and $\hat{N}$ in Eq. (\ref{ABeqQN}), $\hat{H}$ and $\hat{Y}$ in
Eqs. (\ref{Req0}) and (\ref{RneQ})), then Eqs. (\ref{TwoTime}) can be written
in terms of the eigenvalues and eigenfunctions of the system Hamiltonian.
This makes the interpretation of the equations much more physically
transparent. Otherwise (as in Eq. (\ref{RnetQ})), the interpretation is less
obvious. 

In general, all our quantum work theorems are formulated through the
(time-ordered) averages of Hermitian operators along the quantum trajectory.
Thus, the corresponding expressions
are well suited for the evaluation and/or testing by path integral
numerical methods. Finally, the fluctuation theorems for steady-state
currents and charge transport can directly be formulated in terms
of observables, i.e., the probability density distributions of forward and
backward currents.\cite{wil07,gas07a,gas07b,muk07,mae08} This opens up a principal possibility of testing
our Eq. (\ref{JtokQ}), which can also be derived through the fluctuation theorems
for currents.

\section{Conclusion }

We have presented a unified and simple method for generating work theorems 
for classical and quantum Hamiltonian systems, both under
equilibrium conditions and in a steady state. Throughout the paper, we adopt the partitioning of the total Hamiltonian
into the system part, the bath part, and their coupling.  
We have rederived many equalities which are available in the literature 
and obtained a number of new results.  All our expressions can be considered as rigorous mathematical identities, 
because the fulfillment of the Liouville theorem is required only. 

The list of  the work theorems
is not exhausted by those presented in our paper,
and new equalities can easily be generated via Eqs. (\ref{AB2}) and (\ref{AB2Q}), if necessary. Our results
can be useful for obtaining various partition
functions and (generalized) free energies through simulations and/or
measurements performed on nonequilibrium systems. 
The (nonequilibrium) partition functions are important and useful quantities,\cite{zub,gra,her93,pan98,tas06}  
since they can be employed exactly in the same manner as their equilibrium counterparts.  For example, if we differentiate the logarithm of the steady-state partition function $Z_{s0}$ (which corresponds to either classical (\ref{Rne}) or quantum (\ref{RneQ}) steady-state distribution) with respect to the difference of chemical potentials $\Delta_{\mu}$, 
we get the steady-state value of the operator $\hat{Y}(0)$. This operator, which is linearly related to the time-integrated current $\hat{\overline{J}}$, has the meaning of energy we have to spend for establishing the steady-state distribution. 

Finally, we wish to comment on the role of nonconservative forces in establishing and maintaining steady states.  
Deriving our classical (\ref{Rne}) and quantum (\ref{RneQ}) steady-state distributions, we did not invoke any external nonconservative forces for establishing the steady state. The only requirement is that the system under study (either with finite or infinite number of degrees of freedom) is coupled to the bath with infinite number of degrees of freedom, and the thermodynamic limit is applied. \cite{and06} As a result, the system exchanges energy with the Hamiltonian bath during establishing and maintaining the steady state ($\Delta Y$ in Eqs. (\ref{Rne}) and (\ref{RneQ})) but no additional nonconservative dissipative forces are required.  
The distributions (\ref{Rne}) and (\ref{RneQ}) can be derived within the framework of the method of statistical operator by Zubarev \cite{zub} and (generalized version of the) maximum entropy principle by Jaynes.\cite{gra} The use of these distributions and the standard Keldysh Green's functions technique yields the same steady-state averages.\cite{her93,and06,fuj07}   
A controversial question is whether such an approach is adequate for describing quantum transport on the nanoscale,\cite{and06,tod04,div05,gro05,das05} but this is beyond the scope of the present paper.
This picture is in contrast with the steady-state thermodynamics,\cite{pan98,tas06} in which non-conservative dissipative forces are responsible for establishing and maintaining the steady state, and thus an additional "housekeeping heat'' is necessary  to keep the system in the steady state. So, beyond the formal general results, the equalities derived in the present article can be directly compared with their counterparts obtained within framework of the steady-state thermodynamics \cite{sas01,sei05,sei05a,jar06} provided that the "housekeeping heat'' equals to zero. This simply renders the steady-state thermodynamics distributions equilibrium distributions. 
However, the non-conservative forces can straightforwardly be incorporated into our Hamiltonian approach if we switch to the thermostated dynamics.\cite{eva02,wil07}

\begin{acknowledgments}
This work has been supported by NSF-MRSEC DMR0520471 at the University of Maryland and  by the American Chemical Society Petroleum Research Fund (44481-G6).
\end{acknowledgments}

\appendix
\section{Equivalence of Zubarev method of  statistical operator and Keldysh nonequilibrium Green's functions}
Within Zubarev method of statistical operator the steady state average of an operator $G$ is defined as\cite{zub}
\begin{equation}
\overline{G} =  \mbox{Tr} [ \rho(0) G],
\label{zub-aver}
\end{equation}
where 
\begin{equation}
\rho(0) = \lim_{\eta \rightarrow 0} \eta \int_{-\infty}^{0} dt' \exp(\eta t') U(t',0) \rho_{rel} (t') U^{\dagger}(t',0)
\label{zub-dens}
\end{equation}
Here $\rho_{rel}(t)$ is the "relevant statistical distribution" and $U(t,0)$ is the time evolution operator.\cite{zub} We assume that the relevant statistical distribution is  given by the time-independent equilibrium density matrix 
\begin{equation}
\rho_{rel}(t) = \rho(-\infty).
\end{equation} 
We rewrite Eq.(\ref{zub-dens}) in the following form
\begin{equation}
\rho(0) = \lim_{\eta \rightarrow 0} \int_{-\infty}^{0} dt' \frac{d }{dt'} ( \exp(\eta t') ) U(t',0) \rho(-\infty) U^{\dagger}(t',0)
\label{zub-dens2}
\end{equation}
Integrating Eq.(\ref{zub-dens2}) by parts, we obtain
\begin{equation}
\rho(0) =  U(-\infty,0) \rho(-\infty) U^{\dagger}(-\infty,0)
\label{zub-dens3}
\end{equation}
Therefore, the steady-state average value obtained within Zubarev method of statistical operator becomes
\begin{equation}
\overline{G} =  \mbox{Tr} [ U(-\infty,0) \rho(-\infty) U^{\dagger}(-\infty,0) G].
\label{zub-aver2}
\end{equation}
Since the averaging in Keldysh non-equilibrium Green's functions is defined as
$ \mbox{Tr} [ \rho(-\infty)  U^{\dagger}(-\infty,0) G U(-\infty,0) ]$,\cite{keldysh} it is clear that it can be obtained from Zubarev average (\ref{zub-aver2}) by the cyclic permutation of the operators under the trace.


\begin{thebibliography}{20}
\bibitem{foot1}For the steady state to occur, the system should normally be coupled to (at least) two 
different baths which are kept at different chemical potentials. To make the presentation simpler, we employ
the (effective) single-bath description, since our formal expressions can be  straightforwardly generalized  to the multiple baths case.
\bibitem{foot3}If we consider $\rho_{t}$ for any finite $t$, then the classical
and quantum evolution operators are unitary
and thus $Z_{\tau}=Z_{0}$ for any finite system. Since we perform
the thermodynamic limit $N_{B}\rightarrow\infty$, $V_{B}\rightarrow\infty$,
$N_{B}/V_{B}\rightarrow\mathrm{const}$ first and tend $\tau\rightarrow-\infty$
afterwords, then the evolution operators become isometric \cite{tay} and, in general, 
$Z_{\tau}\neq Z_{0}$ (see Refs.\cite{her93,zub,kli07,and06} for more
details).
\bibitem{footN}For example, there exist sounded arguments against attributing
$\Delta_{E}$ the meaning of work in the (standard) quantum Jarzynski
equation, because the so-defined quantity does not properly describe
work fluctuations. \cite{nie05}
\end{thebibliography}
\end{document}